\documentclass[fleqn,twoside]{article}
\usepackage{espcrc2}
\usepackage{graphicx}
\newcommand{\be}{\begin{eqnarray}}
\newcommand{\ee}{\end{eqnarray}}
\newcommand{\ba}{\begin{array}}
\newcommand{\ea}{\end{array}}
\newcommand{\no}{\nonumber}
\newcommand{\la}{\langle}
\newcommand{\ra}{\rangle}

\newcommand{\KEK}
{High Energy Accelerator Research Organization (KEK),
 Tsukuba 305-0801, Japan}
\title{Heavy Quark Physics and Lattice QCD}
\author{Norikazu~Yamada\address{\KEK}}
\begin{document}
 \begin{abstract}
  I review recent progress made on heavy quark physics on the
  lattice.
%
%
 \end{abstract}

\maketitle

\section{INTRODUCTION}

Precise knowledge on $B$ meson decay properties plays an essential
role in testing the unitarity of the CKM matrix
\cite{Cabibbo:1963_yz,Kobayashi:1973_fv}.
The experimental status relevant to the unitarity test is now
promising because experimental uncertainties are already very small or
expected to be reduced well in the near future
\cite{exp_status:2002_ichep}.
On the other hand, theoretical calculations relevant to the test are
still not sufficiently accurate due to the non-perturbative effect of
QCD.
Lattice QCD is an ideal tool to deal with this effect and should be
able to reduce the current theoretical uncertainties
\cite{Kronfeld:2001ss}.

Apart from the asymmetric $e^+e^-$ colliders, there are two important
upcoming experiments for the unitarity test.
The Tevatron Run II will produce a large number of and variety of $b$
and $c$-hadrons, and their basic properties such as mass and lifetime
will be precisely measured \cite{Anikeev:2001rk}.
What is important for us is that once $B^0_s$-$\bar{B}^0_s$ mixing is
observed the mass difference $\Delta M_{B_s}$ will be measured with a
few percent accuracy.
The width difference in the $B_s$ meson system is also expected to be
measured precisely.
Another exciting experiment is the CLEO-c project \cite{Briere:2001rn}.
There, charmed quarkonia, hybrid states and glueballs will be observed.
In particular the leptonic and the semi-leptonic decays of $D$ mesons
are expected to be measured with a few percent level.
Advancing lattice QCD calculations with a view to combining them with
these precise experimental results is an urgent task in front of us.

In this review, I will focus on new and updated calculations of hadron
matrix elements.
The current status and progress made in spectrum calculations and
formulations are not covered, although they are very important and
interesting.
In particular, I will mainly discuss the lattice determination of the
$B^0$-$\bar{B}^0$ mixing amplitude and how to put a strong constraint
on the poorly known CKM element $|V_{td}|$.
The other important quantities such as form factors of semi-leptonic
decays will be mentioned briefly.
Recently several suggestions to improve the limits on accuracy of
present lattice calculations have been made in methodology.
I will briefly introduce some of them before summarizing my point of
view on the current status.

\section{$B^0$-$\bar{B}^0$ mixing}

\subsection{General remarks}

Within the Standard Model, the mass difference of neutral $B$ meson
system is given by
\be
& & \Delta M_{B_q}
 =  \left( \mbox{\rm known factor} \right) \times
    |V_{tb}^*V_{tq}|^2 \no\\
& & \times \frac{\la \bar{B}_q^0|
          \bar{b}\gamma_\mu(1-\gamma_5)q~
          \bar{b}\gamma_\mu(1-\gamma_5)q~
          |B_q^0\ra}
         {M_{B_q}},
\label{eq:B-Bbar_SM_expression}
\ee
where $q$=$d$ or $s$.
$\Delta M_{B_d}$ has already been measured accurately
$\Delta M_{B_d}$ = 0.503 $\pm$ 0.006 ps$^{-1}$ \cite{LEPBOSC:2002},
while for $\Delta M_{B_s}$ only a lower bound is known
($\Delta M_{B_s}$ $>$ 14.4 ps$^{-1}$ at 95 \% CL \cite{LEPBOSC:2002}),
but it is expected to be measured precisely at the Tevatron Run II.
The hadron matrix element in (\ref{eq:B-Bbar_SM_expression}) is
usually parameterized as
\be
 &&   \la\bar{B}_q^0|~
      \bar{b}\gamma_\mu(1-\gamma_5)q~
      \bar{b}\gamma_\mu(1-\gamma_5)q~
      | B_q^0\ra \no\\
 && = \frac{8}{3}f_{B_q}^2 B_{B_q}(\mu_b)M_{B_q}^2,
\ee
where the decay constant $f_{B_q}$ is defined through the following
matrix element,
\be
 &&   \la 0|~\bar{b}\gamma_\mu(1-\gamma_5)q~| B_q^0\ra
    = i f_{B_q} p_\mu.
\ee
Since the four-quark operator receives renormalization, $B_B$ depends
on renormalization scheme and scale.
In the literature, the renormalization scale independent $B$ parameter
is often used, which is defined by
\be
   \displaystyle
   \hat{B}_{B_q}
 = \left[ \alpha_s(\mu_b) \right]^{-\frac{6}{23}}
   \left[ 1 + \frac{\alpha_s(\mu_b)}{4\pi}J_5 \right]
            B_{B_q}(\mu_b),
\ee
where the full expression of $J_{n_f}$ is found in
\cite{Buchalla:1995vs}.
Throughout this review, I quote $\hat{B}_{B_q}$ rather than the scale
dependent one $B_{B_q}(\mu)$.

In the following, I summarize the current status of the quenched and
unquenched calculations of $f_{B_q}$ and $\hat{B}_{B_q}$.

\subsection{Quenched $f_{B_q}$ and $\hat{B}_{B_q}$}

According to the recent Lattice conference reviews
\cite{Onogi:1997kb,Draper:1998ms,Hashimoto:1999bk,Bernard:2000ki,Ryan:2001ej},
the quenched results for $f_{B_d}$ and $\hat{B}_{B_d}$ have been
stable over several years as shown in Table \ref{tab:previous_fB_BB}.
\begin{table}
 \begin{tabular}{l|ll}
  year & $f_{B_d}$ [MeV] & $\hat{B}_{B_d}$ \\ \hline
  '97 Onogi     & 163(12) & \\
  '98 Draper    & 165(20) & 1.32(6)(12)\\
  '99 Hashimoto & 170(20) & 1.23(23)\\
  '00 Bernard   & 175(20) & 1.30(12)\\
  '01 Ryan      & 173(23) & 1.30(12)\\ \hline
 \end{tabular}
 \caption{The quenched $f_{B_d}$ and $\hat{B}_{B_d}$ quoted in the
          recent reviews.
 } 
 \label{tab:previous_fB_BB}
\end{table}
This year JLQCD \cite{Aoki:2002bh} updated their quenched calculation,
which is performed at $\beta$=6.0 using the non-perturbatively $O(a)$
improved Wilson light quark and NRQCD for heavy quarks.
They find $f_{B_d}$ = 158(3)(10) MeV,
$f_{B_s}/f_{B_d}$ = 1.16(1)($^{+04}_{-00}$),
$\hat{B}_{B_d}$ = 1.29(5)(7) and
$\hat{B}_{B_s}/\hat{B}_{B_d}$ = 1.020(24)(15)($^{+4}_{-0}$),
where the first error is statistical, the second systematic and the
third comes from the uncertainty of $m_s$.

\begin{figure}
 \leavevmode
 \includegraphics*[width=7cm,clip]{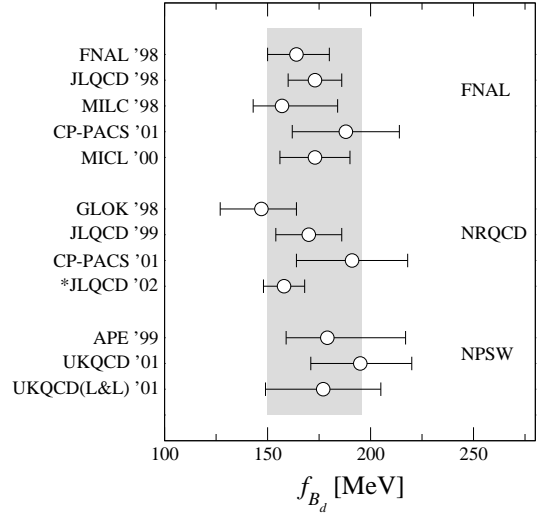}
 \vspace{-3ex}
 \caption{Summary plot of quenched $f_{B_d}$. Shaded band represents
          the current average given by Ryan \cite{Ryan:2001ej}.}
 \label{fig:quench_fB_summary}
\end{figure}

Figure \ref{fig:quench_fB_summary} summarizes recent results for
quenched $f_{B_d}$
\cite{El-Khadra:1997hq,Aoki:1998ji,Bernard:1998xi,AliKhan:2000eg,Bernard:2000ht,AliKhan:1998df,Ishikawa:1999xu,AliKhan:2001jg,Becirevic:1998ua,Bowler:2000xw,Lellouch:2000tw}
including the JLQCD's updated result \cite{Aoki:2002bh}.
The results are grouped into three categories,
``FNAL'' \cite{El-Khadra:1996mp},
``NRQCD'' \cite{Thacker:1990bm} and
``NPSW'', depending on the formulation for heavy quark\footnote{
For a brief summary on the actions used, for example, see
\cite{Ryan:2001ej}.}.
We find an agreement within about 10\% among the results obtained with
the three different actions.
Since the new result by JLQCD is consistent with the previous average
given by Ryan \cite{Ryan:2001ej} (shaded band in the figure),
I keep her values as the summary as of this conference.
\be
 \left.
  \ba{l}
   f_{B_d}^{\rm (N_f=0)} = 173(23) \mbox{MeV},\\
   f_{B_s}^{\rm (N_f=0)} = 200(20) \mbox{MeV},\\
   f_{D_d}^{\rm (N_f=0)} = 203(14) \mbox{MeV},\\
   f_{D_s}^{\rm (N_f=0)} = 230(14) \mbox{MeV}.
  \ea
 \right.
\label{eq:fBD_quench_wa}
\ee
These values are helpful in calibrating new formulation of heavy
quarks.

\begin{figure}
 \leavevmode
 \includegraphics[width=7cm]
 {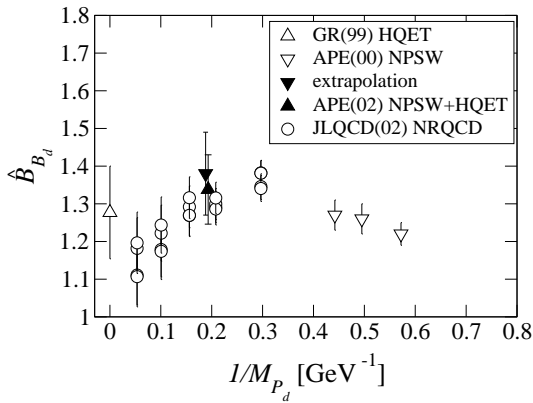}
 \vspace{-3ex}
 \caption{Comparison of $1/M_{P_d}$ dependence of $\hat{B}_{B_d}$.} 
 \label{fig:quench_BB_ape_jlqcd_comparison}
 \vspace{-3ex}
\end{figure}

As for $\hat{B}_{B_q}$, only a couple of calculations are available.
Figure \ref{fig:quench_BB_ape_jlqcd_comparison} shows the $1/M_{P_d}$
dependence of $\hat{B}_{B_d}$ obtained by APE \cite{Becirevic:2000nv}
with the relativistic heavy quark (NPSW) and by JLQCD
\cite{Aoki:2002bh} with NRQCD for heavy quark.
UKQCD has also used the relativistic heavy quark
\cite{Lellouch:2000tw}, but since their result agrees with the APE
result very well it is suppressed in the figure for simplicity.
I also plot the result calculated in the static limit (HQET) by
Gimenez and Martinelli \cite{Gimenez:1996sk}\footnote{
Later their result was corrected in \cite{Gimenez:1998mw}.}.

The $1/M_{P_d}$ dependence of $\hat{B}_{B_d}$ seems inconsistent
between two formulations of heavy quark, but the results agree within
error at the physical point $1/M_{B_d}$=0.19 GeV$^{-1}$.
A possible reason for this disagreement is systematic errors, which
are different for different formulations;
the relativistic action has sizable $O(a^2m^2)$ errors and the NRQCD
approach contains large $O(\alpha_s^2)$ errors in the matching.
In \cite{Becirevic:2001xt}, the results with the relativistic approach
is combined with the one obtained in the static limit and interpolated
to the physical $B_d$ meson mass (filled triangle up).
It is interesting that the result with the combined analysis is
slightly lower than that with extrapolation (filled triangle down) and
the consistency with the NRQCD becomes better.

SPQcdR has started a new calculation at $\beta$=6.45
\cite{Becirevic:2002qr} with the RI-MOM scheme
\cite{Martinelli:1994ty} for renormalization.
The strategy is similar to what APE adopted, namely relying on the
extrapolation in heavy quark mass from around $c$ quark region to the
the physical $b$ quark mass.
However, due to the use of high $\beta$ value, the distance to the
physical $B$ meson mass becomes smaller.
The motivation of this work is to test if the $O(a^2)$ error is under
control.
While the numerical results are still very preliminary, no significant
change has been observed compared to the APE results at $\beta$=6.2
\cite{Becirevic:2000nv}.
This suggests that the discretization error, which is most serious in
this extrapolation method, is under control.

\begin{figure}
 \leavevmode
 \includegraphics[width=7.4cm,clip=]{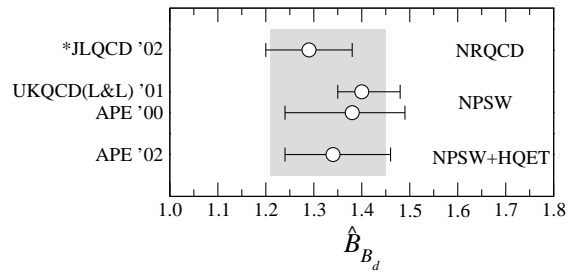}
 \vspace{-4ex}
 \caption{Summary of quenched $\hat{B}_B$.
          Shaded band represents the current average.}
 \label{fig:quench_BB_summary}
\end{figure}

Figure \ref{fig:quench_BB_summary} summarizes the quenched
calculations of $\hat{B}_{B_d}$.
At this moment, I quote the following conservative value as a current
estimate,
\be
 \hat{B}_{B_d}^{\rm (N_f=0)} =1.33(12),
\ee
which is indicated in the figure by shade.

\subsection{Unquenched $f_{B_q}$}
\label{sec:unquenchedfb}

Several large scale unquenched calculations of $f_B$ were carried out
a few years ago, and the recent reviews reported 10--20 \% increase
compared to the quenched ones
\cite{Draper:1998ms,Hashimoto:1999bk,Bernard:2000ki,Ryan:2001ej}.
I wish to discuss that this conclusion needs a reexamination because
of possible effects of chiral logarithms expected from chiral
perturbation theory.

Incorporating heavy quarks into chiral perturbation theory
\textit{via} heavy quark symmetry was first proposed by Wise
\cite{Wise:1992hn} and Burdman and Donoghue \cite{Burdman:1992gh}.
Since then several authors have developed this idea
\cite{Casalbuoni:1996pg}.
Recently Booth \cite{Booth:1995hx} and Sharpe and Zhang
\cite{Sharpe:1996qp} extended the idea to (partially) quenched chiral
perturbation theory ((P)QChPT) to make it applicable to actual lattice
calculations.

In the $N_f$=2 unquenched case, where $u$ and $d$ quarks are treated
dynamically, PQChPT predicts the non-analytic term in $f_{B_d}$ to be
\cite{Sharpe:1996qp}
\be
  - \frac{3}{4}(1+3g^2)~\frac{m_\pi^2}{(4\pi f)^2}
    \ln(m_\pi^2/\Lambda^2)
\label{eq:chiral_log_fBd},
\ee
where $\Lambda$ is the cutoff of the theory and $f$=$f_\pi$ at the
order considered.
While $f_\pi$ has a similar expression for its non-analytic term, in
this case the $B^*B\pi$ coupling $g$, which is only poorly constrained
in $g$=0.3$\sim$0.6 \cite{Casalbuoni:1996pg}, appears because the mass
difference in $B$ and $B^*$ mesons is smaller than the pion
excitation.
For $f_{B_s}$, the expression of the non-analytic term takes a
different form.
In the $N_f$=2 partially quenched case, where $u$ and $d$ quarks are
treated dynamically as before and $s$ appears only as a valence quark,
the prediction for the non-analytic term becomes
\be
 - (1+3g^2)\frac{m_{s\bar{q}}^2}{(4\pi f)^2}
   \ln(m_{s\bar{q}}^2/\Lambda^2)
\label{eq:chiral_log_fBs},
\ee
where
$m_{s\bar{q}}^2=(m_{s\bar{s}}^2+m_\pi^2)/2$.
We observe that while for the chiral behavior of $f_{B_d}$
(\ref{eq:chiral_log_fBd}) indicates significant slope as $m_\pi^2$
vanishes,
(\ref{eq:chiral_log_fBs}) suggests a milder behavior for $f_{B_s}$
because $m_{s\bar{q}}^2$ becomes a constant when
$m_{s\bar{s}}^2\gg m_\pi^2$. 
Lattice data are supposed to show these logarithmic behaviors.

\begin{figure}
 \leavevmode
 \includegraphics[width=7.5cm,clip=]
 {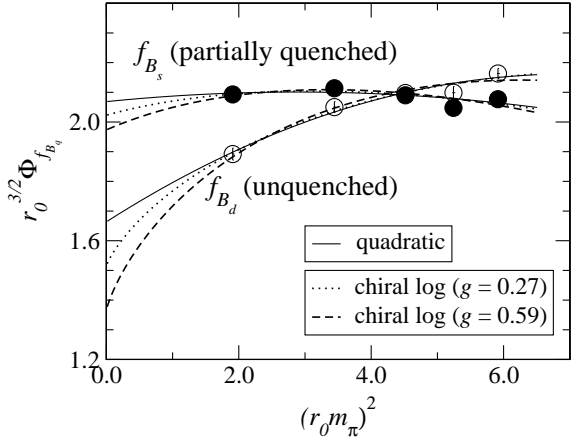}
 \vspace{-6ex}
 \caption{Chiral extrapolation of $\Phi_{f_{B_q}}$ given by
          \cite{Hashimoto:2002vi}.
          Both axis are normalized by Sommer's scale $r_0$
          \cite{Sommer:1993ce}.
 }
 \label{fig:unquench_fsqrt_jlqcd02}
 \vspace{-3ex}
\end{figure}

JLQCD has accumulated twice as many statistics as they had last year
in $N_f$=2 QCD, and updated their analysis of the decay constant
\cite{Hashimoto:2002vi}.
The simulation is performed at $\beta$=5.2 ($a\sim$0.09 fm) with the
NRQCD action for heavy quark and the non-perturbatively
$O(a)$-improved Wilson quark action for both dynamical and valence
light quarks, which range from 0.5 $m_s$ to 1.5 $m_s$.
The chiral extrapolation of $\Phi_{f_{B_q}}$=$f_{B_q}\sqrt{M_{B_q}}$ is
shown in Fig. \ref{fig:unquench_fsqrt_jlqcd02}.
They attempt to fit $\Phi_{f_{B_d}}$ and $\Phi_{f_{B_s}}$ with
(\ref{eq:chiral_log_fBd}) and (\ref{eq:chiral_log_fBs}), respectively,
as well as with a conventional quadratic fit.
They find that $f_{B_d}$ depends on the fit form and the $B^*B\pi$
coupling $g$ significantly, while $f_{B_s}$ does not, in agreement
with the observation made above.
This means that $f_{B_s}$/$f_{B_d}$ has large uncertainties as well.
From this analysis, they give $f_{B_s}/f_{B_d}$=1.24--1.38 as a
preliminary result.

Recently, Kronfeld and Ryan \cite{Kronfeld:2002ab} pointed out the
similar enhancement of the ratio by taking into account the predicted
chiral log and reconsidering the chiral extrapolation.
Their estimate is $f_{B_s}/f_{B_d}$=1.32(8) and
$B_{B_s}/B_{B_d}$=1.00(2).
Both analyses suggest that the ratio can be significantly larger than
the previous world average $f_{B_s}/f_{B_d}$=1.16(5)
\cite{Ryan:2001ej}.

Here let me summarize the $N_f$=2 unquenched calculations of
$f_{B_q}$.
At present, consistency between lattice data and ChPT is not clear
because the value of $g$ is still unknown.
In addition, the $O(1/M)$ contribution to the log term is not known
well \cite{DiBartolomeo:1994ir}.
In this review, the central value of decay constants is taken from the
previous ones \cite{Ryan:2001ej}, which can be considered as those
obtained with a quadratic fit in the chiral extrapolation.
The uncertainty for $f_{B_d}$ associated with the chiral logarithm is
estimated following JLQCD while it is neglected for $f_{B_s}$.
According to their analysis,
\be
  (f_{B_d}^{\rm qua}-f_{B_d}^{\rm ChPT})/f_{B_d}^{\rm qua}
  =0.17,
\ee
where $f_{B_d}^{\rm qua}$ is obtained by quadratic form (as usual) and
$f_{B_d}^{\rm ChPT}$ includes the effect of chiral logarithm.
In obtaining $f_{B_d}^{\rm ChPT}$, they take the upper bound $g$=0.6,
for which the effect becomes maximum.
Taking this uncertainty into account, I quote
\be
 f_{B_d}^{\rm (N_f=2)} &=& 198(30)(^{+~0}_{-34})~~\mbox{MeV},\\
 f_{B_s}^{\rm (N_f=2)} &=& 230(30)~~\mbox{MeV},\\
 f_{B_s}^{\rm (N_f=2)}/f_{B_d}^{\rm (N_f=2)}
                       &=& 1.16(5)(^{+24}_{-0})
\label{unquench_fBsoverfBd},
\ee
as my conservative estimates.

The physical prediction of $f_{B_{d,s}}$ requires three-flavor
dynamical simulations.
MILC has started such an attempt \cite{Bernard:2002ep}.
They employ a highly improved gauge and staggered quark actions
for gauge configurations and the tadpole-improved clover action for
valence quarks, where FNAL formalism is applied for heavy quarks.
The simulation is performed for two lattice spacings $a\sim$ 0.13 and
0.09 fm.
To keep the lattice spacing constant, $\beta$ is adjusted as dynamical
quark mass is varied.
Since $Z_A$ is not available yet, they focus on ratios.
Chiral extrapolations are made in two steps.
First the valence light quark is extrapolated (interpolated) to the
physical up/down (strange) quark mass at each dynamical quark mass.
Then dynamical quark mass is extrapolated to up/down.
The third flavor of dynamical quark is tuned to $m_s$ in advance.

\begin{figure}
 \leavevmode
 \includegraphics[width=7.5cm]
 {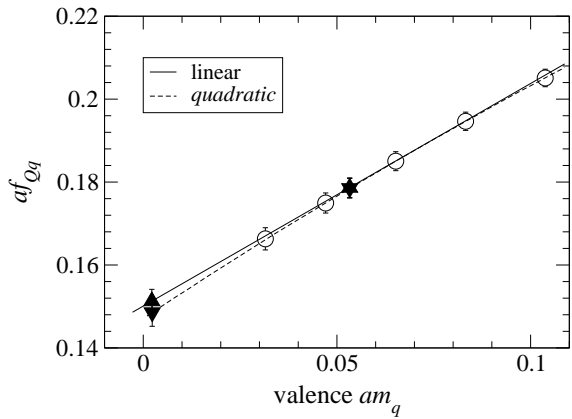}
 \vspace{-5ex}
 \caption{Chiral extrapolation of $af_{Qq}$ in valence quark mass
          given by \cite{Bernard:2002ep}.}
 \label{fig:fsqrt_valence_chiral_milc02}
 \vspace{-3ex}
\end{figure}

Figure \ref{fig:fsqrt_valence_chiral_milc02} shows the chiral
extrapolation of $af_{Qq}$ in valence quark mass.
They attempt to fit the data by two types of fit forms, linear
(solid line) and quadratic (dotted line) functions, and find no
significant deviation between them.
It is interesting to see whether this linear behavior is consistent
with $N_f$=3 PQChPT.
But such a test is complicated because of flavor breaking effects in
pion masses for the staggered quark action \cite{Aubin:2002ss}.

\begin{figure}[t]
 \leavevmode
 \includegraphics[width=7.5cm]{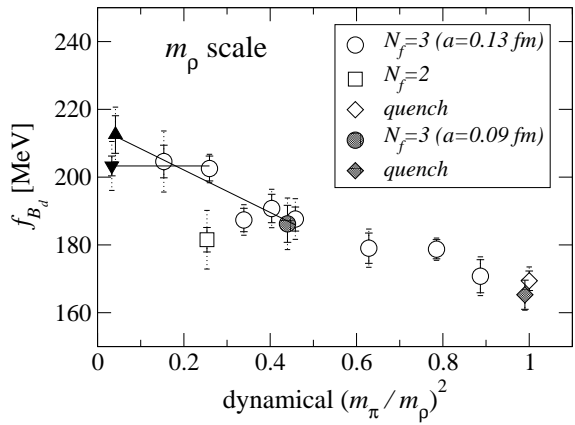}
 \vspace{-5ex}
 \caption{Chiral extrapolation of $f_{B_d}$ in dynamical quark mass
          given by \cite{Bernard:2002ep}.
 }
 \label{fig:fBd_dynamical_chiral_milc02}
\end{figure}

\begin{figure}[t]
 \leavevmode
 \includegraphics[width=7.5cm]{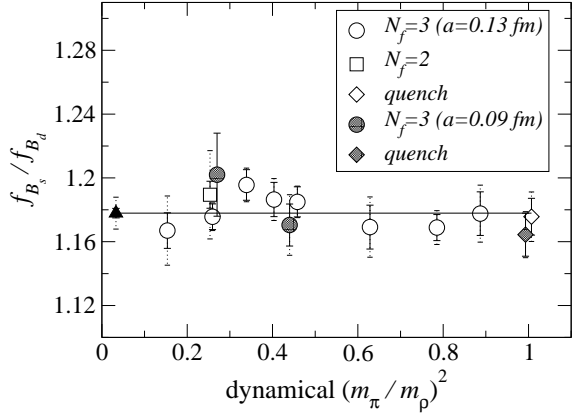}
 \vspace{-5ex}
 \caption{Chiral extrapolation of $f_{B_s}/f_{B_d}$ in dynamical
          quark mass given by \cite{Bernard:2002ep}.
 }
 \label{fig:fBsfBd_dynamical_chiral_milc02}
\end{figure}

The chiral extrapolations of $f_{B_d}$ and $f_{B_s}/f_{B_d}$ in
dynamical quark mass are shown in
Figs. \ref{fig:fBd_dynamical_chiral_milc02} and
\ref{fig:fBsfBd_dynamical_chiral_milc02}, respectively.
In \ref{fig:fBd_dynamical_chiral_milc02}, where the scale is set by
$m_\rho$, a negative slope is seen.
But once one sets the scale by $r_1$ \cite{Bernard:2000gd},
the slope becomes unclear, namely $f_{B_d}$ appears to be independent
on $m_{\rm dyn}$.
In \ref{fig:fBsfBd_dynamical_chiral_milc02}, they find that the ratio
is independent both of $N_f$ and $m_{\rm dyn}$, and quote
$f_{B_s}/f_{B_d}=1.18(1)(^{+4}_{-1})$ (preliminary).
It is very interesting to check if the whole above observations are
consistent with ChPT.

\subsection{unquenched $\hat{B}_{B_q}$}

This year no new or updated calculation on the unquenched
$B$-parameters was presented.
Last year JLQCD reported that the fit form dependence is relatively
mild for $\hat{B}_{B_q}$ in their $N_f=2$ unquenched simulation
\cite{Yamada:2001xp}.
This is consistent with the PQChPT prediction that the chiral
logarithm term for $\hat{B}_{B_d}$ is given by \cite{Sharpe:1996qp} as
\be
 - \frac{1}{2}(1-3g^2)\frac{m_\pi^2}{(4\pi f)^2}
   \ln(m_\pi^2/\Lambda^2)
\label{eq:chiral_log_BBd},
\ee
and hence suppressed compared with the case of $f_{B_d}$ in
(\ref{eq:chiral_log_fBd}), and that chiral logarithm is altogether
absent for $\hat{B}_{B_s}$.
JLQCD also reported that quenching effects are small from a comparison 
with the quenched $\hat{B}_{B_q}$.

\begin{figure}
 \leavevmode
 \includegraphics[width=7.5cm,clip=]{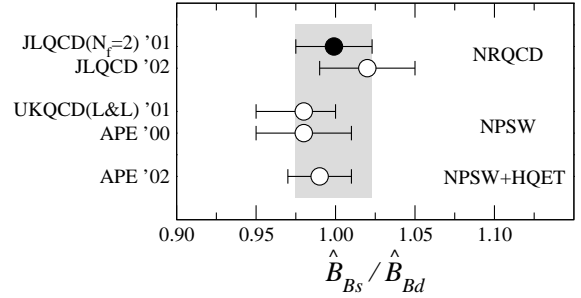}
 \vspace{-6ex}
 \caption{Comparison of quenched and unquenched
   $\hat{B}_{B_s}/\hat{B}_{B_d}$.
   Only the top one is obtained from an unquenched simulation.
   The central value I quote is indicated by shade.}
 \label{fig:unq_BBsBB_summary}
\end{figure}

The ratio $\hat{B}_{B_s}/\hat{B}_{B_d}$ is also stable against the
number of dynamical flavor, the heavy quark action and the fit form in
the chiral extrapolation, and the recent results are compared in
Fig. \ref{fig:unq_BBsBB_summary}, where only the data shown in top
is unquenched results.

Since at present there is only one unquenched calculation by JLQCD, I
take their values as current estimates, and quote
\be
 \hat{B}_{B_d}^{\rm (N_f=2)} &=& 1.34(13),\\
 \frac{\hat{B}_{B_s}^{\rm (N_f=2)}}{\hat{B}_{B_d}^{\rm (N_f=2)}}
                             &=&0.999(24)
\label{unquench_BBsoverBBd}.
\ee
It should be noted that the ratio is already very precise.

\subsection{SU(3) flavor breaking ratio}

The Tevatron Run II experiment is expected to yield a precise value
for the mass difference of $B_s$ meson system, $\Delta M_s$, in the
near future \cite{Anikeev:2001rk}.
Once $\Delta M_s$ is determined precisely, assuming the three
generation unitarity ($|V_{ts}|\approx |V_{cb}|$) we can put a strong
constraint on the poorly known CKM matrix element $|V_{td}|$ through
the SU(3) flavor breaking ratio of the hadron matrix element,
\be
     \frac{\Delta M_{B_s}}{\Delta M_{B_d}}
 &=& \frac{|V_{ts}|^2}{|V_{td}|^2}~
     \frac{M_{B_s}}{M_{B_d}}~\xi^2,
 \label{eq:SU(3)FVR}
\ee
where $\xi=(f_{B_s}\sqrt{\hat{B}_{B_s}})/(f_{B_d}\sqrt{\hat{B}_{B_d}})$.

Until recently it has been expected that $\xi$ can be easily
determined within a few percent accuracy by lattice QCD.
However, the presence of chiral logarithm makes it more difficult.
As we discussed in Sec.~\ref{sec:unquenchedfb}, $f_{B_s}/f_{B_d}$ can
still have a large systematic uncertainty while the ratio of
$\hat{B}_{B_q}$ does not.
Following the values given in (\ref{unquench_fBsoverfBd}) and
(\ref{unquench_BBsoverBBd}), I quote as my personal estimate,
\be
 \xi^{\rm (N_f=2)} &=& 1.16(6)(^{+24}_{-0}).
\ee
Kronfeld and Ryan also gave a similar value ($\xi$=1.32(10)) in their
analysis \cite{Kronfeld:2002ab}.

I now discuss that a precise determination of $|V_{td}|$ is still
possible if one considers the Grinstein ratio of decay constants
\cite{Grinstein:1993ys} defined by
\be
 R_1=(f_{B_s}/f_{B_d})/(f_{D_s}/f_{D_d}).
\ee
Rewriting (\ref{eq:SU(3)FVR}) in terms of $R_1$, one obtains
\be
& &       \left(\frac{\Delta M_{B_s}}{\Delta M_{B_d}}
          \right)_{\rm Run II}
 =  \frac{|V_{ts}|^2}{|V_{td}|^2}
    \left(\frac{M_{B_s}}{M_{B_d}}\right)_{\rm known}\no\\
& & \hspace{4ex}\times
         \left(\frac{f_{D_s}^2}{f_{D_d}^2}\right)_{\rm CLEO-c}
          \left(\frac{\hat{B}_{B_s}}{\hat{B}_{B_d}}~
                R_1^2 \right)_{\rm Lattice}.
\ee
As indicated in the subscript, the ratio $f_{D_s}/f_{D_d}$ is expected
to be measured in CLEO-c, and $\hat{B}_{B_s}/\hat{B}_{B_d}$ can be
determined by lattice QCD precisely.
Therefore $R_1$ is the only remaining quantity to be fixed.

We expect this task is achievable in lattice QCD since variations
under chiral and heavy quark expansions will both cancel out in the
ratio to a large extent, leaving only a small deviation from unity.

This year JLQCD presented their preliminary result of $R_1$
\cite{Onogi:2002_lat02}.
\begin{figure}[t]
 \leavevmode
 \includegraphics[width=7.5cm,clip=]{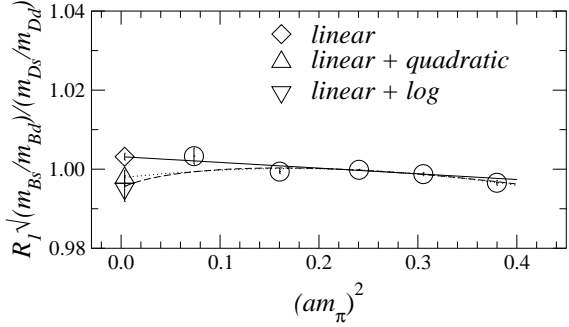}
 \vspace{-6ex}
 \caption{Chiral extrapolation of
    $R_1\times\sqrt{(M_{B_s}/M_{B_d})/(M_{D_s}/M_{D_d})}$ given by
    \cite{Onogi:2002_lat02}.
 }
 \label{fig:JLQCD_GR_onogi}
\end{figure}
The simulation is carried out on $N_f$=2 dynamical configurations at
$\beta$=5.2.
In order to handle the $c$ quark, FNAL formalism is applied for heavy
quarks.
Figure \ref{fig:JLQCD_GR_onogi} shows the chiral behavior of
$R_1\times\sqrt{(M_{B_s}/M_{B_d})/(M_{D_s}/M_{D_d})}$.
Several fit forms are attempted as shown in the figure.
They found that the result is relatively insensitive to the fit form
and its dependence is, at most, about 1\%.
Their preliminary result is $R_1 = 1.018(06)(10)$, where the first
error is statistical and the second represents the systematic
uncertainty.

\begin{figure}[t]
 \leavevmode
  \includegraphics[width=6.5cm,clip=]{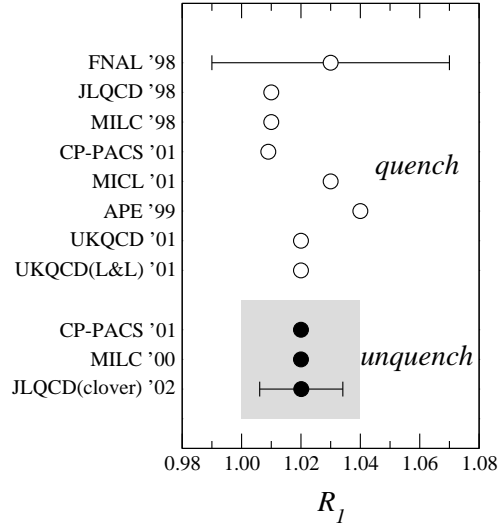}
 \vspace{-6ex}
 \caption{Summary of the Grinstein ratio.
 }
 \label{fig:summary_GR}
\end{figure}

In Fig. \ref{fig:summary_GR}, I gather the Grinstein ratio obtained by
previous calculations.
Only FNAL'98 \cite{El-Khadra:1997hq} and JLQCD'02(clover)
\cite{Onogi:2002_lat02} explicitly calculated $R_1$.
For other results, I only show their central values.
It can be seen that this ratio is very stable and there is only a
negligible fluctuation.
At present, I quote
\be
  R_1^{\rm (N_f=2)} = 1.02(2).
\ee
From a comparison with quenched and unquenched ($N_f$=2) results, it
is unlikely that the central value and its accuracy change drastically
when one goes to $N_f$=3.

As we have seen, once $\Delta M_{B_s}$, $f_{D_s}$ and $f_{D_d}$ are
measured in the forthcoming experiments, we can test the unitarity at
a few percent level through $|V_{td}|$.
To gain more confidence, the current values of $R_1$ and
$\hat{B}_{B_s}/\hat{B}_{B_d}$ should be confirmed by several groups.

\section{Form factors}

\subsection{heavy to heavy transition}

In order to test the CKM unitarity, a precise determination of
$|V_{cb}|$ is indispensable, as it sets the normalization of the
parameters ($\rho$,$\eta$).
To this end, $\bar{B}\rightarrow D^*l\bar{\nu_l}$ is most promising.
The differential decay rate is given by
\be
   \frac{d\Gamma}{d\omega}
 = \left(\mbox{known factor}\right)\ |V_{cb}|^2\
   |{\cal F}_{B\rightarrow D^*}(\omega)|^2,
\ee
where $\omega=v'_{D^*}\cdot v_B$.
At zero recoil ($\omega=1$),
${\cal F}_{B\rightarrow D^*}(1)=h_{A_1}(1)$, where $h_{A_1}(1)$ is
given by
\be
  \la D^*(v)|\bar{c}\gamma_\mu\gamma_5 b |\bar{B}(v)\ra
= i\sqrt{2m_B2m_{D^*}}\epsilon'_\mu h_{A_1}(1).
\ee
Last year Fermilab group demonstrated a precise determination of
${\cal F}_{B\rightarrow D^*}(1)$ by a full use of heavy quark symmetry
and obtained
${\cal F}_{B\rightarrow D^*}(1)
 =0.913{^{+0.0238}_{-0.0173}}{^{+0.0171}_{-0.0302}}$
in the quenched approximation \cite{Hashimoto:2001nb}.
In my opinion the lattice method for calculation of this decay mode
has been established.
The extension to unquenched calculations remains to be done.
It is worth noting that chiral logarithm terms  would not affect this
accuracy because they are suppressed by $(1/M)^2$.

At this conference a new calculation of
$\Lambda_b\rightarrow\Lambda_c l\nu$ was presented by Tamhankar
\cite{Tamhankar:2002_lat02}.
This process is experimentally challenging, but gives an independent
determination of $|V_{cb}|$.
They use the quenched 20$^3\times$64 lattice with $a^{-1}$=1.32 GeV.
The actions used are $O(a^2,\alpha_sa^2)$ improved gauge and tadpole
improved clover and FNAL approach is taken for $b$ and $c$ quarks.
The chiral limit is not taken at present.
Instead, they investigate the initial heavy quark mass ($m_b$)
dependence with fixed $m_c$ and $m_{\rm light}$.
No clear mass dependence has been observed in their preliminary
analysis.

\subsection{heavy to light transition}

\begin{figure}
 \leavevmode
 \includegraphics[width=7cm,clip=]{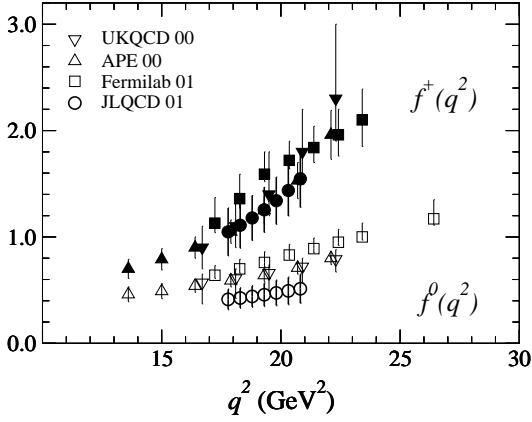}
 \vspace{-3ex}
 \caption{The current status of quenched calculations of the
          $B\rightarrow\pi l\nu$ form factors.
 }
 \label{fig:b2pi_by_four_group}
\end{figure}

Once $|V_{td}|/|V_{cb}|$ and angle $\phi_1$($\beta$) are determined
precisely enough, the location of the apex is essentially fixed in the
$\rho$-$\eta$ plane.
Then the next step is to test the consistency of the CKM mechanism,
for example, by measuring other CKM elements such as $|V_{ub}|$.
$B\rightarrow \pi l\nu$ is one of the simplest choice for such a
purpose.
The definition of form factors, $f^+$ and $f^0$, is given by
\be
& &    \la \pi(p')\ |\ \bar{q}\gamma_\mu b\ |\ B(p)\ra\no\\
&=&    (p_\mu + p'_\mu - \frac{m_B^2-m_\pi^2}{q^2}q_\mu)
       f^+(q^2)\no\\
& & +\ \frac{m_B^2-m_\pi^2}{q^2}q_\mu f^0(q^2),
\ee
where $q=p-p'$.
Figure \ref{fig:b2pi_by_four_group} \cite{Hashimoto:2002_CKMworkshop}
shows the quenched calculations made by four major groups
\cite{Bowler:1999xn,Abada:2000ty,El-Khadra:2001rv,Aoki:2001rd}.
Data from different groups agree with each other, but only within a
large uncertainty ($\sim$20\%).
Unquenched simulations remain to be done.
Moreover the issue of chiral logarithm could be a significant source
of uncertainty in this case.
More studies are needed before finalizing the form factor calculation.

This year SPQcdR made two contributions on the heavy to light vector
meson semi-leptonic decays,
$B\rightarrow \rho l\nu$ ($D\rightarrow K^* l\nu$ )
\cite{Abada:2002ie} and
$B\rightarrow K^*\gamma$ \cite{Becirevic:2002_B2K*gamma}.
Both are performed at $\beta$=6.2 and 6.45 with non-perturbatively
$O(a)$-improved Wilson action for light and heavy quarks, and the
currents are also improved non-perturbatively.
The form factors of these decays obtained around $c$ quark mass is
extrapolated to the physical $b$ quark mass by using the HQET scaling
laws for a fixed value of $v\cdot p'$ or $q^2$=0.
According to their preliminary results, no scaling violation has been
seen for both matrix elements, but the statistical uncertainty is
still significant.
At present, chiral extrapolations are carried out by assuming the form
factors to be a linear function of light vector meson mass.
But in the future a guiding principle will be necessary to make a
extrapolation more reliable.

\subsection{Determination of $g_{D^*D\pi}$ coupling}

The determination of the coupling constant $g$ defined in ChPT, or
equivalently $g_{P^*P\pi}$ given by
\be
&&    \la P(p)\ \pi(q)\ |\ P^*(p',\lambda)\ra\no\\
&=&\!\!\!\!
    - g_{P^*P\pi}(q^2) q\cdot\epsilon^\lambda(p')
      (2\pi)^4\delta(p'-p-q),\\
&&    g_{P^*P\pi}
  =   \lim_{q^2\rightarrow m_\pi^2}g_{P^*P\pi}(q^2).
\ee
is now extremely important because this coupling plays a crucial role
in the chiral extrapolation of decay constant and in the normalization
of one of the $P\rightarrow\pi$ form factor $f^+(q^2)$ at
$q^2$=$q_{\rm max}^2$.
$g_{P^*P\pi}$ is directly calculable on the lattice, and more
interestingly $g_{D^*D\pi}$ has been measured in CLEO through the
$D^{*+}\rightarrow D^{0(+)}\pi^{+(0)}$ decays \cite{Ahmed:2001xc}
while $B^*\rightarrow B\pi$ is prohibited kinematically.
The relation between $g$ and $g_{P^*P\pi}$ is given by ChPT as
$g_{P^*P\pi}=2\ g\ \sqrt{m_{P^*}m_P}/f_\pi$ up to $O(1/M)$ and
$O(m_\pi^2)$ corrections.
Using an LSZ reduction of the pion and the PCAC relation, the
calculation of above matrix element reduces to that of
\be
      \la P(p)\ |\ \bar{q}\gamma_\mu\gamma_5 q\ |\ P^*(p+q)\ra,
\ee
namely, to the calculation of the form factors describing the above
matrix element.

\begin{figure}
 \leavevmode
 \includegraphics[width=8cm]
 {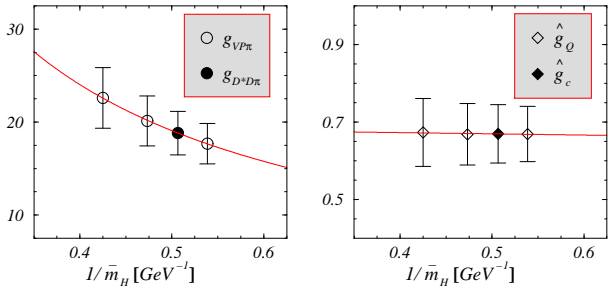}
 \vspace{-6ex}
 \caption{The $1/M_P$ dependence of $g_{D^*D\pi}$ and $g$ in the
          chiral limit.
 }
 \label{fig:D*2Dpi_Herdoiza}
\end{figure}

The first exploratory study to determine $g_{P^*P\pi}$ on the lattice
was made in the static limit by UKQCD \cite{deDivitiis:1998kj}.
This year the determination of $g_{D^*D\pi}$ at physical $D$ meson
mass was presented by Herdoiza in the quenched approximation
\cite{Abada:2002xe}.
Figure \ref{fig:D*2Dpi_Herdoiza} shows the $1/M_P$ dependence of
$g_{D^*D\pi}$ and $g$ in the chiral limit.
It is found that the $1/M_P$ dependence of the former is
completely canceled by the factor $\sqrt{m_P^*m_P}$.
After interpolation to the physical $D$ meson mass, they obtain
$g_{D^*D\pi}$=18.8(2.3)($^{+1.1}_{-2.0}$), which agrees with
$g_{D^*D\pi}$=17.9(0.3)(1.9) measured by CLEO \cite{Ahmed:2001xc}.

It seems that the first step toward the determination has been
completed.
The next step is the unquenched simulation with special care to the
chiral extrapolation.

\section{Improvements in methodology}

\subsection{KS as light quark}

We have seen that the issue associated with chiral logarithm is
present almost everywhere.
One possible course to simulate light sea quark masses is to employ
staggered fermions.

At the last conference, Wingate \textit{et al.} performed an
exploratory study of $B$ mesons with the NRQCD and staggered actions
on a coarse lattice ($a^{-1}$=0.8 GeV), and pointed out a difficulty
associated with contaminations from unphysical modes
\cite{Wingate:2001mp}.
This year they repeated the similar calculations on slightly finer
lattices using $N_f$=0 ($a^{-1}$=1 GeV) and $N_f$=3 ($a^{-1}$=1.3 GeV)
configurations \cite{Wingate:2002ry}.
They found that such unphysical modes can be essentially removed by
the use of finer lattice.  They also observed that the correlation
functions with both parities is well separated by applying suitable
fit forms and the constrained curve fitting \cite{Lepage:2001ym}.

\begin{figure}
 \vspace{-10ex} 
 \leavevmode
 \includegraphics[width=7.5cm,clip=on]
 {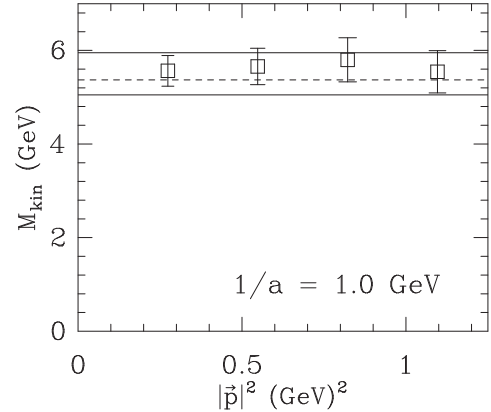}
 \vspace{-6ex} 
 \caption{Comparison of the mass obtained from dispersion relation
          (open square) with that using perturbation theory
          (dashed line).
          The solid lines indicate the perturbative uncertainty in the
          latter.
 }
 \label{fig:Wingate}
 \vspace{-3ex} 
\end{figure}

They calculated $B_s$ meson spectrum, and made a comparison of
$M_{B_s}$ obtained from dispersion relation with that using
perturbation theory as shown in Fig. \ref{fig:Wingate}.
These two definitions of mass agree quite well.
They also obtained
$f_{B_s}^{\rm (N_f=0)}$=225$\pm$9(stat.)$\pm$20(pert.) MeV
in the quenched approximation and this value is reasonably consistent
with the current quenched world average
$f_{B_s}^{\rm (N_f=0)}$=200(20) MeV.
More importantly they conclude that the unquenching dramatically
improves the hyperfine splitting $M_{B_s^*}-M_{B_s}$ and results in
42.5$\pm$3.7 MeV.
These observations indicate that there is no practical problem in the
combination of NRQCD and staggered fermions.

\subsection{Anisotropic lattice}

\begin{figure}
 \vspace{-3ex} 
 \leavevmode
  \includegraphics[width=7.5cm,clip=on]
  {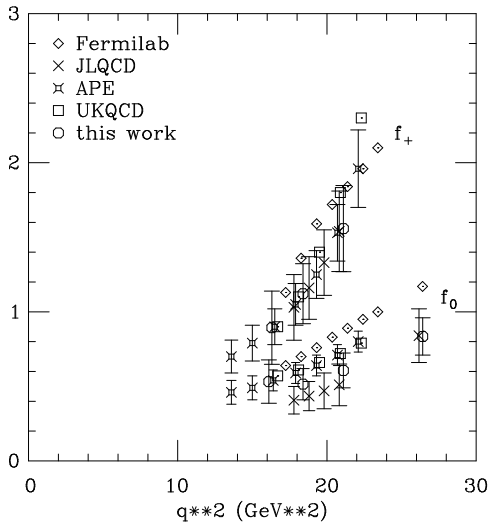}
 \vspace{-6ex} 
 \caption{Comparison of form factors obtained with iso- and
          aniso-tropic lattice given by \cite{Shigemitsu:2002wh}.
 }
 \label{fig:Shigemitsu}
 \vspace{-3ex} 
\end{figure}

The motivation for the use of anisotropic lattices in heavy-light
system is to obtain clean signals in the form factor calculations.
The method has been applied to the $B\rightarrow\pi$ semileptonic
decay by Shigemitsu \textit{et al.} \cite{Shigemitsu:2002wh}.
The simulation is performed on a $12^3\times 48$ anisotropic quenched
lattice with $a_s/a_t$=2.71 ($1/a_s$=1.2GeV) using the
tadpole-improved Symanzik gluon action, NRQCD for heavy quarks and
D234 action for light quarks.
They apply the constrained fits \cite{Lepage:2001ym} to the
three-point functions to obtain the form factors shown in
Fig. \ref{fig:Shigemitsu}.
Their results agree well with those with isotropic lattices.
In spite of a relatively small number of configurations (199),
the statistical error in this work is comparable or even smaller than
those of JLQCD, which used NRQCD heavy quarks and was obtained with
more than 1,000 configurations \cite{Aoki:2001rd}.

\subsection{Step scaling method}

At this conference a new method to calculate $f_B$ in non-perturbative
accuracy was presented by Petronzio \cite{Guagnelli:2002jd}, based on
a non-perturbative recursive finite size technique.
The explicit calculation of $f_B$ is made in the Schr\"odinger
functional setup with several quenched lattices.
Most of relevant renormalization constants are obtained
non-perturbatively.
My understanding of this method is as follows.
Let us start from the following identity,
 \be
&&\hspace*{-2ex}
    f_B(12\mbox{ GeV})|_{1.6}
 =  f_B(12\mbox{ GeV})|_{0.4}\no\\
&&\hspace*{2ex}
    \times
    \frac{f_B(12\mbox{ GeV})|_{0.8}}
         {f_B(12\mbox{ GeV})|_{0.4}}
    \times
    \frac{f_B(12\mbox{ GeV})|_{1.6}}
         {f_B(12\mbox{ GeV})|_{0.8}}
\label{eq:finitesize_fB_1},
\ee
where $f_B(12 \mbox{ GeV})|_L$ denotes $f_B$ measured on the lattice
with $a^{-1}$=12 GeV and the physical volume $L$ fm.
The first factor $f_B(12\mbox{ GeV})|_{0.4}$ is interpreted as $f_B$
on a finite volume and the remaining factors are considered as
corrections to obtain $f_B$ with large volume.
It should be noted that with this lattice spacing the $b$ quark can be
suitably treated by the relativistic formalism and that the above
equation holds even if the smallest lattice is in the deconfined phase.
Since it is difficult to realize the lattice simulation with both
$a^{-1}$=12 GeV and L=1.6 fm simultaneously, consider the following,
\be
&&\hspace*{-2ex}
    f_B(12\mbox{ GeV})|_{1.6}
\approx
    f_B(12\mbox{ GeV})|_{0.4}\no\\
&&\hspace*{2ex}
    \times
    \frac{f_B(6\mbox{ GeV})|_{0.8}}
         {f_B(6\mbox{ GeV})|_{0.4}}
    \times
    \frac{f_B(3\mbox{ GeV})|_{1.6}}
         {f_B(3\mbox{ GeV})|_{0.8}}
\label{eq:finitesize_fB_2},
\ee
which is obtained by replacing the second and third factors by those
obtained on coarser lattices but with the same physical volume.
They performed above calculation and obtained $f_B$=170(11)(5)(22) MeV
and $f_{B_s}$=192(9)(5)(24), where the first error is statistical and
the second and third are systematic.
These values agree well with the current quenched estimates
(\ref{eq:fBD_quench_wa}).

The questions are how (\ref{eq:finitesize_fB_2}) is justified,
what is $f_B$ in the deconfined phase, and what is the advantage of
this method compared to the calculation on a single lattice with
$a^{-1}$=3 GeV and 1.6 fm.
The possible systematic uncertainties are still unclear to me.
The same calculation in the static limit would be helpful to make this
clear.

\section{Summary}

We are now going into an exciting era because Belle, BaBar, Tevatron
Run II and CLEO-c will be giving us a wealth of precision data for $B$
and $D$ mesons soon.

One of the CKM matrix elements that these experiments should pin down
is $|V_{td}|$.
An important point realized recently is that the amplitude of
$B^0$-$\bar{B}^0$ mixing necessary to convert the experimental mass
difference to this matrix element suffers from a sizable uncertainty
of 10--20\% due to chiral logarithms in the chiral extrapolation.
However, one can avoid this problem if one introduces the Grinstein
ratio of decay constants.
It is now possible to determine the relevant quantities on the lattice
as accurately as 5\% and better, and the precise determination of
$|V_{td}|$ is realized when $\Delta M_{B_s}$, $f_{D_d}$ and $f_{D_s}$
are measured in the forthcoming experiments.

For $|V_{cb}|$, needed for the normalization of the matrix elements,
the situation is more promising because the form factor of the
$B\rightarrow D^*l\nu$ decay can be determined on the lattice with a
few percent accuracy.
The lattice method for this calculation is established, and does not
receive significant effects from chiral logarithms.
It only remains to apply the method to unquenched calculations, and
independent checks using other processes are in progress.

The extraction of $|V_{ub}|$ from $B\rightarrow\pi(\rho)$ is still
theoretically challenging as it is so in experiment.
The form factor calculations still have large uncertainties
($\sim$20\%) even in the quenched approximation.
Further improvements are necessary.
In particular, it is essential to find out how one can secure clean
signals from simulations and how to overcome the problem of chiral
logarithms.

To achieve better accuracy, improvements in the methodology are also
crucial.
For the issue of chiral logarithm, lowering the quark mass provides
the direct route for resolution for which staggered light quarks may
be helpful. Clean signals in form factor calculations are obtained
with anisotropic lattices.
Finally the non-perturbative accuracy might be achieved by the
development of the step scaling method.

\section*{Acknowledgment}

I would like to thank
D.~Becirevic,
C.~Bernard,
S.~Gottlieb,
S.~Hashimoto,
G.~Herdoiza,
T.~Kaneko,
A.~Kronfeld,
L.~Lellouch,
T.~Onogi,
R.~Petronzio,
J.~Reyes,
J.~Shigemitsu,
A.~Soni,
S.~Tamhankar
A.~Ukawa and
M.~Wingate
for useful discussion and comments.
This work is supported by the JSPS Research Fellowship.



\begin{thebibliography}{99}

\bibitem{Cabibbo:1963_yz}
  N.~Cabibbo,
  Phys.\ Rev.\ Lett.\  {\bf 10} (1963) 531.

\bibitem{Kobayashi:1973_fv}
  M.~Kobayashi and T.~Maskawa,
  Prog.\ Theor.\ Phys.\  {\bf 49} (1973) 652.

\bibitem{exp_status:2002_ichep}
  For experimental status, for example, see,
  A. Stocchi,
  talk at ICHEP 2002, Amsterdam, 24-31 July, 2002,
  http://www.ichep02.nl/.

\bibitem{Kronfeld:2001ss}
  For a recent review, for example, see,
  A.~S.~Kronfeld,
  talk at 9th International Symposium on Heavy Flavor Physics,
  Pasadena, California, 10-13 Sep 2001,
  hep-ph/0111376;
%
  L.~Lellouch,
  talk at ICHEP 2002, Amsterdam, 24-31 July, 2002,
  http://www.ichep02.nl/.

\bibitem{Anikeev:2001rk}
  K.~Anikeev {\it et al.},
  hep-ph/0201071.

\bibitem{Briere:2001rn}
  R.~Galik, these proceedings;
  R.~A.~Briere {\it et al.},
  CLNS-01-1742.

\bibitem{LEPBOSC:2002}
  LEP B oscillations working group,
  http://lepbosc.web.cern.ch/LEPBOSC/.

\bibitem{Buchalla:1995vs}
  G.~Buchalla, A.~J.~Buras and M.~E.~Lautenbacher,
  Rev.\ Mod.\ Phys.\  {\bf 68} (1996) 1125
  and references therein.

\bibitem{Onogi:1997kb}
  T.~Onogi,
  Nucl.\ Phys.\ Proc.\ Suppl.\  {\bf 63} (1998) 59.

\bibitem{Draper:1998ms}
  T.~Draper,
  Nucl.\ Phys.\ Proc.\ Suppl.\  {\bf 73} (1999) 43.

\bibitem{Hashimoto:1999bk}
  S.~Hashimoto,
  Nucl.\ Phys.\ Proc.\ Suppl.\  {\bf 83} (2000) 3.

\bibitem{Bernard:2000ki}
  C.~W.~Bernard,
  Nucl.\ Phys.\ Proc.\ Suppl.\  {\bf 94} (2001) 159.

\bibitem{Ryan:2001ej}
  S.~M.~Ryan,
  Nucl.\ Phys.\ Proc.\ Suppl.\  {\bf 106} (2002) 86.

\bibitem{Aoki:2002bh}
  S.~Aoki {\it et al.}  [JLQCD Collaboration],
  hep-lat/0208038;
  in preparation;
  N.~Yamada {\it et al.}  [JLQCD Collaboration],
  Nucl.\ Phys.\ B (Proc.\ Suppl.)\ 94 (2001) 379.

\bibitem{El-Khadra:1997hq}
  A.~X.~El-Khadra {\it et al.},
  Phys.\ Rev.\ D {\bf 58} (1998) 014506.

\bibitem{Aoki:1998ji}
  S.~Aoki {\it et al.}  [JLQCD Collaboration],
  Phys.\ Rev.\ Lett.\  {\bf 80} (1998) 5711.

\bibitem{Bernard:1998xi}
  C.~W.~Bernard {\it et al.},
  Phys.\ Rev.\ Lett.\  {\bf 81} (1998) 4812.

\bibitem{AliKhan:2000eg}
  A.~Ali Khan {\it et al.}  [CP-PACS Collaboration],
  Phys.\ Rev.\ D {\bf 64} (2001) 034505.

\bibitem{Bernard:2000ht}
  C.~W.~Bernard {\it et al.} [MILC Collaboration],
  Nucl.\ Phys.\ Proc.\ Suppl.\  {\bf 94} (2001) 346;
%
 hep-lat/0206016.

\bibitem{AliKhan:1998df}
  A.~Ali Khan {\it et al.},
  Phys.\ Lett.\ B {\bf 427} (1998) 132.

\bibitem{Ishikawa:1999xu}
  K.~I.~Ishikawa {\it et al.}  [JLQCD Collaboration],
  Phys.\ Rev.\ D {\bf 61} (2000) 074501.

\bibitem{AliKhan:2001jg}
  A.~Ali Khan {\it et al.}  [CP-PACS Collaboration],
  Phys.\ Rev.\ D {\bf 64} (2001) 054504.

\bibitem{Becirevic:1998ua}
  D.~Becirevic \textit{et al.},
  Phys.\ Rev.\ D {\bf 60} (1999) 074501.

\bibitem{Bowler:2000xw}
  K.~C.~Bowler \textit{et al.}
  [UKQCD Collaboration],
  Nucl.\ Phys.\ B {\bf 619} (2001) 507.

\bibitem{Lellouch:2000tw}
  L.~Lellouch and C.~J.~Lin [UKQCD Collaboration],
  Phys.\ Rev.\ D {\bf 64} (2001) 094501.

\bibitem{El-Khadra:1996mp}
  A.~X.~El-Khadra, A.~S.~Kronfeld and P.~B.~Mackenzie,
  Phys.\ Rev.\ D {\bf 55} (1997) 3933.

\bibitem{Thacker:1990bm}
  B.~A.~Thacker and G.~P.~Lepage,
  Phys.\ Rev.\ D {\bf 43} (1991) 196;
%
  G.~P.~Lepage \textit{et al.},
  Phys.\ Rev.\ D {\bf 46} (1992) 4052.

\bibitem{Becirevic:2000nv}
  D.~Becirevic \textit{et al.}
  Nucl.\ Phys.\ B {\bf 618} (2001) 241.

\bibitem{Gimenez:1996sk}
  V.~Gimenez and G.~Martinelli,
  Phys.\ Lett.\ B {\bf 398} (1997) 135.

\bibitem{Gimenez:1998mw}
  V.~Gimenez and J.~Reyes,
  Nucl.\ Phys.\ B {\bf 545} (1999) 576.

\bibitem{Becirevic:2001xt}
  D.~Becirevic \textit{et al.},
  JHEP {\bf 0204} (2002) 025.

\bibitem{Becirevic:2002qr}
  D.~Becirevic \textit{et al.}
  [SPQcdR collaboration],
   these proceedings,
   hep-lat/0209131.

\bibitem{Martinelli:1994ty}
  G.~Martinelli \textit{et al.},
  Nucl.\ Phys.\ B {\bf 445} (1995) 81;
  A.~Donini \textit{et al.},
  Eur.\ Phys.\ J.\ C {\bf 10} (1999) 121.

\bibitem{Wise:1992hn}
  M.~B.~Wise,
  Phys.\ Rev.\ D {\bf 45} (1992) 2188.

\bibitem{Burdman:1992gh}
  G.~Burdman and J.~F.~Donoghue,
  Phys.\ Lett.\ B {\bf 280} (1992) 287.

\bibitem{Casalbuoni:1996pg}
  For a comprehensive review, for example, see,
  R.~Casalbuoni \textit{et al.},
  Phys.\ Rept.\  {\bf 281} (1997) 145.

\bibitem{Booth:1995hx}
  M.~J.~Booth,
  Phys.\ Rev.\ D 51 (1995) 2338.

\bibitem{Sharpe:1996qp}
  S.~R.~Sharpe and Y.~Zhang,
  Phys.\ Rev.\ D 53 (1996) 5125.

\bibitem{Hashimoto:2002vi}
  S.~Hashimoto {\it et al.} [JLQCD Collaboration],
  these proceedings,
  hep-lat/0209091.

\bibitem{Sommer:1993ce}
  R.~Sommer,
  Nucl.\ Phys.\ B {\bf 411} (1994) 839

\bibitem{Kronfeld:2002ab}
  A.~S.~Kronfeld and S.~M.~Ryan,
  Phys.\ Lett.\ B {\bf 543} (2002) 59;
%
  these proceedings,
  hep-lat/0209083.

\bibitem{DiBartolomeo:1994ir}
  For the determination of the $P^*P\pi$ coupling taking into account
  the $1/M_P$ corrections, see, for example,
  N.~Di Bartolomeo \textit{et al.},
  Phys.\ Lett.\ B {\bf 347} (1995) 405.

\bibitem{Bernard:2002ep}
  C.~Bernard {\it et al.}  [MILC Collaboration],
  these proceedings,
  hep-lat/0209163.

\bibitem{Aubin:2002ss}
  C.~Aubin {\it et al.},
  these proceedings,
  hep-lat/0209066,
  and references therein.

\bibitem{Bernard:2000gd}
  C.~W.~Bernard {\it et al.},
  Phys.\ Rev.\ D {\bf 62} (2000) 034503.

\bibitem{Yamada:2001xp}
  N.~Yamada {\it et al.}  [JLQCD Collaboration],
  Nucl.\ Phys.\ Proc.\ Suppl.\  {\bf 106} (2002) 397.

\bibitem{Grinstein:1993ys}
  B.~Grinstein,
  Phys.\ Rev.\ Lett.\  {\bf 71} (1993) 3067.

\bibitem{Onogi:2002_lat02}
  T.~Onogi \textit{et al.} [JLQCD Collaboration],
  these proceedings.

\bibitem{Hashimoto:2001nb}
  S.~Hashimoto \textit{et al.},
  Phys.\ Rev.\ D {\bf 66} (2002) 014503.

\bibitem{Tamhankar:2002_lat02}
  S.~Tamhankar and S.~Gottlieb [MILC Collaboration],
  these proceedings.

\bibitem{Hashimoto:2002_CKMworkshop}
  This figure was presented by S.~Hashimoto in his talk given at
  Workshop on the CKM Unitarity triangle,
  CERN, Geneva, February 13-16th, 2002,
  http://ckm-workshop.web.cern.ch/.

\bibitem{Bowler:1999xn}
  K.~C.~Bowler {\it et al.}  [UKQCD Collaboration],
  Phys.\ Lett.\ B {\bf 486} (2000) 111.

\bibitem{Abada:2000ty}
  A.~Abada \textit{et al.},
  Nucl.\ Phys.\ B {\bf 619} (2001) 565.

\bibitem{El-Khadra:2001rv}
  A.~X.~El-Khadra \textit{et al.},
  Phys.\ Rev.\ D {\bf 64} (2001) 014502.

\bibitem{Aoki:2001rd}
  S.~Aoki {\it et al.}  [JLQCD Collaboration],
  Phys.\ Rev.\ D {\bf 64} (2001) 114505.

\bibitem{Abada:2002ie}
  A.~Abada \textit{et al.}
  [SPQcdR collaboration],
   hep-lat/0209116.

\bibitem{Becirevic:2002_B2K*gamma}
  D.~Becirevic [SPQcdR collaboration],
  talk at ICHEP 2002,
  http://www.ichep02.nl/.

\bibitem{Ahmed:2001xc}
  S.~Ahmed {\it et al.}  [CLEO Collaboration],
  Phys.\ Rev.\ Lett.\  {\bf 87} (2001) 251801;
  A.~Anastassov {\it et al.}  [CLEO Collaboration],
  Phys.\ Rev.\ D {\bf 65} (2002) 032003.

\bibitem{deDivitiis:1998kj}
  G.~M.~de Divitiis \textit{et al.}
  [UKQCD Collaboration],
  JHEP {\bf 9810} (1998) 010.

\bibitem{Abada:2002xe}
  A.~Abada {\it et al.},
   hep-ph/0206237;
%
  these proceedings,
  hep-lat/0209092.

\bibitem{Wingate:2001mp}
  M.~Wingate, J.~Shigemitsu and G.~P.~Lepage,
  Nucl.\ Phys.\ Proc.\ Suppl.\  {\bf 106} (2002) 379.

\bibitem{Wingate:2002ry}
  M.~Wingate \textit{et al.},
  these proceedings,
  hep-lat/0209096.

\bibitem{Lepage:2001ym}
  G.~P.~Lepage \textit{et al.},
  Nucl.\ Phys.\ Proc.\ Suppl.\  {\bf 106} (2002) 12.

\bibitem{Shigemitsu:2002wh}
  J.~Shigemitsu \textit{et al.},
   hep-lat/0207011;
%
  these proceedings,
  hep-lat/0208062.

\bibitem{Guagnelli:2002jd}
  M.~Guagnelli \textit{et al.},
   hep-lat/0206023;
%
  these proceedings,
  hep-lat/0209113.

\end{thebibliography}
\end{document}